\documentclass{article}
\usepackage{tikz}

\PassOptionsToPackage{numbers, compress}{natbib}
\usepackage[dblblindworkshop,final]{neurips_2025}

\workshoptitle{AI for Music}
\usepackage{natbib}
\usepackage[utf8]{inputenc} % allow utf-8 input
\usepackage[T1]{fontenc}    % use 8-bit T1 fonts
\usepackage{hyperref}       % hyperlinks
\usepackage{url}            % simple URL typesetting
\usepackage{booktabs}       % professional-quality tables
\usepackage{amsfonts}       % blackboard math symbols
\usepackage{nicefrac}       % compact symbols for 1/2, etc.
\usepackage{microtype}      % microtypography
\usepackage{xcolor}         % colors
\title{Who Gets Heard? Rethinking Fairness in AI for Music Systems}
\author{
\textbf{Atharva Mehta}$^{1}$\thanks{Equal Contribution} \quad
\textbf{Shivam Chauhan}$^{1}\footnotemark[1]$ \quad
\textbf{Megha Sharma}$^{1}$ \quad
\textbf{Gus Xia}$^{1}$ \\
\textbf{Kaustuv Kanti Ganguli}$^{2}$ \quad
\textbf{Nishanth Chandran}$^{3}$ \quad
\textbf{Zeerak Talat}$^{4}$ \quad
\textbf{Monojit Choudhury}$^{1}$ \\
$^{1}$MBZUAI, UAE \quad
$^{2}$Zayed University, UAE \quad
$^{3}$Microsoft, India \quad
$^{4}$University of Edinburgh, UK \\[4pt]
\texttt{\{atharva.mehta, shivam.chauhan, megha.sharma\}@mbzuai.ac.ae} \\
\texttt{\{gus.xia, monojit.choudhury\}@mbzuai.ac.ae}\\ 
\texttt{kaustuvkanti.ganguli@zu.ac.ae} \quad
\texttt{nichandr@microsoft.com} \quad
\texttt{z@zeerak.org} \\[4pt]
}

\begin{document}

\maketitle

\begin{abstract}
In recent years, the music research community has examined risks of AI models for music, with generative AI models in particular, raised concerns about copyright, deepfakes, and transparency. In our work, we raise concerns about cultural and genre biases in AI for music systems (music-AI systems) which affect stakeholders—including \textit{creators}, \textit{distributors}, and \textit{listeners}—shaping representation in AI for music. These biases can misrepresent marginalized traditions, especially from the Global South, producing inauthentic outputs (e.g., distorted ragas) that reduces \textit{creators'} trust on these systems. Such harms risk reinforcing biases, limiting creativity, and contributing to cultural erasure. To address this, we offer recommendations at dataset, model and interface level in music-AI systems.
\end{abstract}

\section{Introduction}
AI models are rapidly transforming the music landscape with generative AI models that generate indistinguishable music compared to human-composed music\cite{malik2017neural}. State-of-the-art generative music models have improved significantly in automatic evaluation metrics and human evaluations \cite{agostinelli2023musicgen, melechovsky2024mustango};  further bringing the generated music samples closer to human-composed music. The increasing capability of generative systems to produce music across diverse genres and perform strongly in human evaluations has motivated the development of advanced composition tools that integrate generative AI models. These advances enable new forms of expression, providing a new interface that can lower access barriers, e.g., allowing a form of ``vibe music composition'' that allows people to compose music without technical skills with instruments or in-depth knowledge of music~\cite{herremans2017functional}.
Due to the ease of composition, researchers have recently raised concerns about the use of generative AI models for music generation (generative music models)~\cite[e.g.,][]{10.1145/3600211.3604686, sag2023copyright, de2025towards, mccormack2019autonomy, seaver2022computing, morgan2023computing}, highlighting risks such as copyright infringement, undetected deep-fakes,
market saturation, legal compliance, and nonadherence to human requests and preferences. Rising concerns among researchers and artists have led to concerted efforts to strengthen the safety of generative music models, addressing issues such as AI-generated music detection\cite{jimaging11070242, rahman2024sonics}, deepfake detection\cite{afchar2024detecting}, and audio watermarking\cite{epple2024watermarking, san2025latent, liu2025xattnmark}.
However, from the perspective of fairness (as defined previously in NLP \cite{Choudhury_Deshpande_2021}), very limited research has examined representational bias and cultural inclusion in music datasets and generative models. Similar concerns arose earlier in NLP: \citet{bender2021dangers} showed that larger datasets do not ensure diversity, often overrepresenting harmful views such as misogyny and white supremacy, while \citet{joshi-etal-2020-state} highlighted disparities in language inclusion and resources. These findings led to the creation of bias detection benchmarks \cite{10.1162/coli_a_00524} and responsible AI frameworks \cite{bai2022constitutional} to mitigate stereotypes and systemic inequities.
Similarly, for music, \citet{mehta2024missing, mehta2025music} shows a huge skew in global representation\footnote{Throughout the remainder of the paper, we use ``bias'' and ``representational bias with respect to data and system output representation interchangeably.,} with a decreasing focus on Global South cultures, including South Asia, South East Asia, Oceania, and the Middle East, as well as genres like country \& folk music, based on different definitions for fairness. Their analysis draws attention to the potential implications of such systemic biases, including widening economic disparities, limited global creativity in music, and, eventually, cultural erosion on a global scale. To fully address these implications, it is crucial to recognize that such biases are not accidental or harmless; they arise from long-standing patterns of cultural dominance, market-focused choices, and platforms that favor certain types of music while leaving others out \cite{wang2024fairness}. Adding to this complexity is the long-tailed and evolving nature of music itself. 
Musical genres are fluid, hybrid, and constantly emerging in local, diaspora, and transnational contexts \cite{holt2019genre,connell2004world}. It is practically and epistemologically impossible for any AI system to equally represent and capture the entirety of global musical expressions. 
Although these studies are crucial in understanding the metalevel effects of biases in music-AI systems, they do not capture the direct consequences of such biases on the diverse stakeholders in the music ecosystem, including \textit{creators, distributors, platforms}, and \textit{listeners}. With the integration of AI into composition and distribution pipelines, the roles and performance of each stakeholder are being reconfigured. 

Thus, we examine the impact of biased representation of music in AI systems on different stakeholders in the music-AI ecosystem and discuss potential interventions at the data and interface levels to address misrepresentation and underrepresentation of musical traditions. Building on this, our goal is to guide future research at the intersection of generative AI, creativity, and social responsibility by addressing the following questions: (RQ1) Who are the stakeholders in the music-AI ecosystem and what are implications of representational biases on the stakeholders? (RQ2) What design choices and technical challenges in music-AI systems shape how representational biases emerge and affect stakeholders? (RQ3) Given the long tail of genres and the evolving nature of music, what system-, dataset- and model-level strategies can ensure fairness when equal representation is impractical? 

We begin by identifying key stakeholders in the music-AI ecosystem, including \textit{composers, listeners, distributors, teachers}, and \textit{students}. We conducted unstructured interviews~\cite{zhang2009unstructured} with researchers, practitioners, and professionals in music and AI to understand risks associated with representational bias and fairness, particularly around cultural inclusion, misrepresentation, and their implications for stakeholders. Based on these discussions, we analyze each implication's impact on every stakeholder. In addition, to mitigate biases in music-AI systems, we propose a set of recommendations and open research questions for music \& AI community. 
We situate music-AI within broader discussions of AI ethics and fairness, arguing that development must move beyond efficiency to support diverse and equitable creative expression.

\section{RQ1 (i): Who are the stakeholders?}
Stakeholders in music have long shaped the way music is produced, distributed, and experienced. Although their core roles remain, \textit{composers}, \textit{listener}s, \textit{distributors}, \textit{educators}, and \textit{students}, their practices and interactions continue to evolve with technological change. Even though a single person in the music and AI ecosystem has the flexibility of performing multiple roles, we focus solely on the core roles themselves assuming that each person performs a single role. 
\textbf{Creators (composers and performers)}
\textit{Creators (composers and performers) }are the ones who write a musical piece in any form. At the individual level, this includes bands, DJs, and singer-songwriters; at the group level, cultural communities and traditions that preserve music as identity and heritage. To assist composers in compositions, AI is integrated into Digital Audio Workstations (DAWs), enabling creative ideas, effectively editing music with minimal effort, and enhancing live performances by providing accompaniment to artists in various forms, overall reducing the cost of music production. 
\textbf{Distributors and marketing professionals}
\textit{Distributors} and marketing professionals bridge \textit{creators} and \textit{listeners} (discussed in detail later). They can facilitate audience targeting for \textit{creators }and enhance \textit{listener} experience through preference-based distribution and marketing. Individually, they may be freelance promoters, curators, or bloggers; collectively, they include record labels, streaming platforms (\href{https://open.spotify.com/}{Spotify}), and large promotion networks (\href{https://www.soundon.global/?lang=en}{SoundOn by TikTok}). Distributors and marketing professionals make use of recommendation systems based on Non-generative AI algorithms enabling \textit{listener}-personalized recommendations, and market analysis, while generative models can create promotional materials such as short clips or videos tailored to the target market.
\textbf{Listeners}
\textit{Listeners} are those who listen to music, often with varying degrees of attention and engagement. \textit{Listeners} experience and evaluate music both individually (e.g., through apps, study playlists, or background use) and collectively (fan communities, concert audiences). Non-generative AI recommendation systems aim to personalize and improve music discovery such as \href{https://www.spotify.com/us/discover/weekly/}{Spotify’s Discover Weekly}, helping \textit{listeners} with playlist generation and context-aware recommendations.
\textbf{Teachers and experts}
\textit{Teachers and experts} (music educators, scholars, and practitioners) play a vital role in preserving and transmitting musical knowledge. Individually, they mentor \textit{students} and guide practice; at the group level, institutions and musicological communities curate archives, curricula, and research such as \href{https://www.ram.ac.uk/}{The Royal Academy of Music}. AI tools can improve pedagogy and analysis, facilitate genre and music structure discovery and evolution, and provide new insights into human creativity and artistic expression.
\textbf{Students}
\textit{Students} are learners who engage with music through formal or informal education, contributing to the preservation and evolution of cultural and creative traditions. At the individual level, they may study instruments, practice specific genres, or experiment with new forms. At the collective level, they form ensembles, learning communities, or cultural cohorts that sustain musical knowledge across generations. AI systems can support \textit{students} by providing feedback on performance (eg, \href{https://www.tonara.com/}{Tonara}), generating accompaniment, offering interactive learning tools, and simulating performance environments.

\section{RQ1 (ii): Implications of representation bias on stakeholders} \label{sec:socio-cultural}
Now that we understand the different stakeholders in the music-AI ecosystem, we can examine the social, cultural, and ethical issues that arise when these groups are overlooked \cite{smuha2021beyond,hooker2021moving}.
\textbf{Misrepresentation Harms}
Misrepresentation of culturally specific or underrepresented music can distort productions: \textit{creators} face a mischaracterization that undermines authenticity\footnote{Expression of the society's historically important musical traditions, repertoires, and practices.} and trust further undermining artist performances; \textit{listeners} intend to explore diversity, but the available content misrepresents it, shaping their perception incorrectly; and \textit{teachers} and \textit{students} can unknowingly propagate these distortions in education if they use such tools for creating educational content\cite[e.g.,][]{blodgett2017racial,koenecke2020racial,zhang2012vision,huang2025visbias,hamidieh2024identifying}.
\textbf{Homogenization and Exposure Bias}
Homogenization and exposure bias compound the problem. Probabilistic models produce uniform outputs shaped by dominant datasets, and recommendation systems reinforce \textit{listener} exposure to familiar music \cite{yamada2022traditional,kumar2014identifying}. \textit{Creators} struggle to innovate or express cultural depth; small-scale \textit{distributors} and labels who produce and distribute underrepresented music are marginalized; \textit{listeners} develop narrow preferences, limiting discovery; and \textit{students}, who need exposure to diverse musical genres, limit their creative discovery process. 
\textbf{Cultural Erosion \& Widening Economic Disparities}
Generative models can erode cultural identity by homogenizing traditional music, reducing access to distinctive regional forms \cite{smith2021decolonizing,burnett2023music,ardalan2025ai}. Due to lack of music discovery often the under-represented genres and practitioners of such genres face limited attention from \textit{listeners}; \textit{distributors} may prioritize popular genres over traditional ones, further drawing these genres to a position of disadvantage;  eventually widening the economic disparity between \textit{creators} composing popular and under-represented genres. This leads to a snowball effect wherein lesser \textit{creators} from the community pursue these genres as career due to lack of financial resources, eventually which can lead to erosion of such genres. \textit{Listeners} encounter barriers to engaging with authentic musical cultures, leaving scarce options for music discovery, and \textit{students} risk learning culturally homogenized material.
\textbf{Opaque Training Processes and Datasets}
Closed source models like \href{https://www.udio.com/}{Udio} and many open source models including MusicLM\cite{agostinelli2023musiclm}, etc are trained on closed datasets often not accessible to the community. This leads to training models on uncurated datasets with unclear composition and intent \cite{gray2019ghost}. Opaque datasets and models make it impossible to trace where music originates. As a result, AI-generated music can mimic styles or compositions without proper attribution, and human \textit{creators} lose recognition for their work. As a result, \textit{listeners}, \textit{teachers}, and \textit{students} may unknowingly treat AI-generated content as authentic, further eroding appreciation of human creativity.

\section{RQ2: Design choices \& Technical Challenges}
\label{sec:methodgaps}

The ethical implications discussed above, such as cultural bias and misrepresentation, often stem from technical gaps embedded in generative systems that stem from how data, representations, and interfaces are designed. Although these appear system-centered, they ultimately shape the socio-cultural gaps as discussed in Section \ref{sec:socio-cultural}, shaping the kind of music people can generate, access, and experience. \textbf{First}, genre labels in datasets and interfaces are often flattened into ambiguous categories such as world or ethnic, collapsing rich traditions into catch-alls and reinforcing cultural erasure. This reflects a Western preoccupation with classification, where hyperspecific microgenres (e.g., melodic house) coexist alongside the homogenization of non-Western forms, which are treated as undifferentiated. \textbf{Second}, training data rarely retains regional or cultural metadata (e.g., place of origin, instruments, performance context, or linguistic association), making it impossible for models to distinguish between culturally rooted genres and generic patterns. \textbf{Third}, symbolic formats like MIDI, while compact and portable, encode Western tonal and rhythmic defaults, marginalizing microtonality, heterophony, and non-metric rhythms that fall outside its representational frame. \textbf{Fourth}, prompt interfaces often guide users through culturally narrow defaults, such as mood and genre tags such as cinematic or relaxing, which, combined with overreliance on English metadata and tags, further restrict exploratory access to underrepresented traditions. Together, these design choices produce a structurally homogenized music space, where non-Western and fluid traditions are not only underrepresented but actively constrained by the infrastructures of generation.

\section{RQ3: Recommendations}
To address technical shortcomings in trained music-AI systems, we propose recommendations across three levels—datasets, models, and system/interfaces—aimed at improving transparency and bridging the technical gaps that contribute to socio-cultural inequities in music-AI.
\textbf{Dataset level}. Music datasets should be better documented following practices like datasheets and data statements \cite{boyd2021datasheets,bender2021dangers} including origin, recording conditions, \textit{artists/creators} involved (to account for attribution), instruments, genres, regions, style of play and permissions. Data sets should be audited for all genres to estimate the representation of different genres similar to \citet{mehta2024missing} or using entropy to measure genre diversity and supported with richer labels rather than vague tags like "world music". Where sensitive or sacred recordings are involved, communities should be consulted, consent obtained, and material labeled to prevent casual misuse.
\textbf{Model level}. Models should support traceability, letting users see what data influenced an output, which aids transparency, credit, and user trust. Training methods must reduce overfitting to dominant genres (e.g., reweighting, balanced sampling) and include parameters to explore underrepresented styles. Evaluation should move beyond signal quality to account for genre diversity, style preservation, and cultural variation, even if such measures are still complex and emerging.
\textbf{System/interface level}. Interfaces shape user access: genre and region controls should be simple, visible, and not hidden under “advanced settings.” Systems should avoid flattening labels like world music, instead surfacing specific terms (e.g., Gnawa, Taiko) and presenting contextual information about style, region, or instruments. Tool tips or links can guide users toward learning about traditions they generate, fostering awareness and inclusion.
\textbf{Governance level.} Fairness requires structural measures: involving musicians and cultural \textit{teachers and experts} in system design, defining guidelines for use of sensitive music, enabling opt-out mechanisms for \textit{creators}, and supporting compensation or licensing models where appropriate. Policymakers should extend the regulation of AI to include cultural expression and creative labor.

\section{Conclusion \& Future Work}
In this paper, we identified key stakeholders in the music-AI ecosystem and examined the implications of representational bias for each group. We highlighted socio-cultural and technical gaps that further enforce under-representation in music-AI systems and proposed recommendations across datasets, models, and interfaces. Looking ahead, several directions of future research open up: evaluating fairness and cultural representation in generated music; ensuring credit and consent for \textit{creators (composers)} and communities with appropriate policies and governance; supporting cultural fidelity without reinforcing stereotypes; extending symbolic infrastructures like MIDI to non-Western traditions; and addressing language-driven biases in prompts and metadata. Together, these directions chart a path toward music-AI systems that are not only technically capable but also more inclusive and culturally respectful.

\bibliographystyle{plainnat}
\bibliography{neurips_2025}

\end{document}